\documentclass[twocolumn,pra,superscriptaddress]{revtex4-2}

\usepackage{epsfig}
\usepackage{graphicx}
\usepackage{amsmath}
\usepackage{amssymb}
\usepackage{epstopdf}
\usepackage{xcolor}
\usepackage{bm}
\usepackage{dsfont}
\usepackage{hyperref}
\usepackage{braket}
\usepackage{tikz}
\usepackage[normalem]{ulem}

\usepackage{subcaption}

\begin{document}

\title{Deterministic preparation of optical qubits with coherent feedback control}
\author{Amy Rouillard}
\affiliation{School of Chemistry and Physics, University of KwaZulu-Natal, Durban, South Africa}
\author{Tanita Permaul}
\affiliation{School of Chemistry and Physics, University of KwaZulu-Natal, Durban, South Africa}
\author{Sandeep K. Goyal}
\email{skgoyal@iisermohali.ac.in}
\affiliation{Department of Physical Sciences,
	Indian
	Institute of Science Education \&
	Research (IISER) Mohali, Sector 81 SAS Nagar,
	Manauli PO 140306 Punjab India.}

\author{Thomas Konrad}
\email{konradt@ukzn.ac.za}
\affiliation{School of Chemistry and Physics, University of KwaZulu-Natal, Durban, South Africa}
\affiliation{National Institute of Theoretical and Computational Sciences (NITheCS), KwaZulu-Natal, South Africa}

\begin{abstract}
We propose a class of preparation schemes for orbital angular  momentum and polarisation qubits carried by single photons or classical states of light based on coherent feedback control by an ancillary degree of freedom of light.  The preparation methods use linear optics and  include the transcription of an arbitrary polarisation state onto a two-level OAM system (swap) for arbitrary OAM values $\pm\ell$ within a light beam, i.e. without spatial interferometer.  The preparations can be carried out with unit efficiency independent from the potentially unknown initial state of the system. Moreover, we show how to translate measurement-based qubit control channels into coherent feedback schemes for optical implementation. 
\end{abstract}

\maketitle
\section{Introduction}

The control of quantum systems is an important prerequisite for all quantum technologies, i.e.\ quantum computation and communication~\cite{Nielsen2011} as well as quantum metrology~\cite{Haroche2011, Wineland2011}. Closed-loop quantum control techniques, probing the quantum system, can be classified in two groups, measurement-based techniques~\cite{WisemanBook} and coherent-feedback techniques~\cite{Lloyd00, Nelson_etal00, HiroseCappellaro16, konrad2020robust}. For quantum communication, which involves non-classical states of light, measurement based-control is technically challenging since non-invasive measurements that do not absorb photons require coupling to matter or other photons, which is in general weak and difficult to realise~\cite{Haroche2011} or comes with low efficiency~\cite{wang2015quantum}.    
Therefore coherent control feedback techniques have the potential to improve quantum communication and all quantum technologies using light as a medium.  

Here, we give a recipe to translate any measurement-based control channel for two-level systems into a coherent feedback control scheme that can be implemented with optics given that the control channel is described by two Kraus operators. Our scheme can be generalised to systems with Hilbert spaces of finite dimension and arbitrary number of Kraus operators.

Moreover, we present a class of control methods based on coherent feedback where a particular degree of freedom of light can be controlled by a second degree of freedom. 
Such methods enable, e.g., the deterministic preparation of any pure polarisation state in a Mach-Zehnder interferometer for a photon with initially mixed polarisation.  Similarly, any coherent superposition  $\alpha \ket{-\ell}+ \beta \ket{\ell}$ can be prepared from an unknown possibly incoherent mixture of light modes with orbital angular momentum (OAM) values $\pm \ell$ by means of a spatial interferometer (path-degree of freedom) or, using polarisation in a single light beam. Unlike other methods of preparation for spatial modes, e.g. by means of spatial light modulation, the present one suggests low noise and works ideally without photon loss. For the special case of equally weighted superpositions,  $\alpha=\beta=1/\sqrt{2}$,  we show how to produce such a target state for all pairs of OAM values $\pm \ell $, simultaneously.

All preparation schemes described below work equally well for single-photon and classical states of light including thermal states with finite temperature. The underlying coherent feedback control method is robust and can be used to protect a target state against noise \cite{konrad2020robust}.

The conditions required for a control channel to successfully drive an arbitrary initial state into a desired target state are reviewed in Section~\ref{Sec:SFP}.  The optical implementation scheme, for various degrees of freedom of light, is informed by a decomposition technique~\cite{sandeep_csd} described in Section~\ref{Sec:CS}. In Section~\ref{Sec:2-out} this technique is applied to unitary couplings that realise coherent feedback control, and the translation from measurement-based to coherent feedback control is presented. Section~\ref{Sec:2-out} also details the optical realisation of the coherent feedback control of polarisation and OAM qubits. Methods which allow for the repeated application of coherent feedback control are discussed in Section~\ref{Sec:Reuse}. Such repetitions also protect target states against noise \cite{konrad2020robust} and can steer the system into target dynamics \cite{Uys2018SFP}. In Section~\ref{Sec:Examples} we give three examples of coherent feedback control and their respective optical implementations. A discussion of the results in Section~\ref{discussion} concludes this article.


\section{Coherent feedback control}\label{Sec:SFP}
In order to prepare light in a given target state $\ket{T}\in\mathcal{H}_s$, where $\mathcal{H}_s$ is the Hilbert space of the system, the following theorem can be applied \cite{konrad2020robust}.  Consider a trace-preserving quantum channel  \$ described by an $n$-element set of Kraus operators $K_i$ satisfying the fix-point condition
\begin{align}
K_i \ket{T} = z_i\ket{T} \label{eq:fp_contition}
\end{align}
for all $i=0,\dots, n-1$ with $z_i\in \mathbb{C}$. In addition, let the Kraus operators obey a second condition,
\begin{align}
\operatorname{span}\{K_i^\dagger \ket{T}\}_{i=0,\dots,n-1} = \mathcal{H}_s\, . \label{eq:span_contition}
\end{align}
Then {an arbitrary initial} state of the system $\rho$ converges to target state $\ket{T}\in\mathcal{H}_s$ under repeated application of the channel $\$$, 
 \begin{align}
\rho \rightarrow \$(\rho) = \sum_{i=0}^{n-1} K_i \rho K_i^\dagger\, . \label{Eq:channel_k}
 \end{align}
The second condition~\eqref{eq:span_contition} ensures that the system is driven towards the target state \cite{konrad2020robust},  while fix point condition~\eqref{eq:fp_contition} arrests it there. Various construction methods of control channels are discussed in \cite{ konrad2020robust,Uys2018SFP}.

Any such channel $\$$ can be implemented by a unitary time evolution $U$ which couples the system to a suitable ancilla system (quantum controller) in initial state $\ket{0}$, such that
 \begin{align}
 \$(\rho) = \operatorname{Tr}[U(\ket{0}\bra{0} \otimes \rho) U^\dagger]   \,.\label{Eq:channel_u}
 \end{align}
The quantum controller probes the system's state and accentuates coherent feedback accordingly. 
If we compare Eqs.~\eqref{Eq:channel_k} and \eqref{Eq:channel_u} the Kraus operators are revealed as function of the time evolution,  
\begin{align}
K_i = \braket{i|U|0}\,, \label{Eq:K_U}
\end{align} 
where $(\ket{i})_{i=1,\ldots n}$ is an orthonormal basis of the quantum controller's Hilbert space. Obviously, there is no way to map all initial states of system and controller onto a single state by a unitary (i.e.\ reversible) evolution.  Consequently, the information about the arbitrary, and possibly unknown, initial state of the system must be transferred to the controller system during their {coupling~\cite{samal2011experimental}}. This implies that after one application of the channel $\$$, the controller will be in an unknown state and must be reset to $\ket{0}$ if $\$$ is to be repeated using the same controller. 

The resetting of the controller presents the main challenge in the repeated implementation of the control scheme and will be discussed in Section~\ref{Sec:Reuse} (abbreviated to Sec.~\ref{Sec:Reuse}). It is possible to design a control channel that reaches the target within a single application, an example of such a channel is given in Sec.~\ref{Sec:Basic}. However, it is still useful to investigate methods which allow for multiple iteration of the channel as this will protect the state against noise.


\section{Cosine-Sine decomposition}\label{Sec:CS}
The time evolution $U$ can be {constructed} from elementary single- and two-partite unitary gates. For this purpose, we employ the Cosine-Sine (CS) decomposition of an arbitrary $U$ into a product of conditional unitaries  and Hadamard gates. When applied to light, these conditional unitaries correspond to operations on a single degree of freedom of light ({system}) conditioned on the state of another ancillary degree of freedom (controller).

An arbitrary  $(m+n)\times(m+n)$ unitary matrix $U_{m+n}$ ($n\geq m$) can be decomposed into $n\times n$ and $m\times m$ unitaries, $L_n, R_n$ and $L_m, R_m$ respectively, as well as a cosine-sine (CS) matrix, according to the CS decomposition~\cite{sandeep_csd}, given by
\begin{equation}
U_{m+n} = \begin{pmatrix}
L_{m} & 0\\
0 & L'_n
\end{pmatrix}(\mathcal{S}_{2m}\oplus\mathds{1}_{n-m})\begin{pmatrix}
R^\dagger_{m} & 0\\
0 & R'^{\dagger}_n
\end{pmatrix}\,,
\end{equation}
where $\mathds{1}_{n-m}$  is the identity matrix in $(n-m)$-dimensions and the so-called CS matrix $\mathcal{S}_{2m}$ reads
\begin{equation}\label{Eq:CS} 
\mathcal{S}_{2m} = \begin{pmatrix}
C_m & S_m\\
-S_m & C_m
\end{pmatrix}
\end{equation}
with {$C_m \equiv \sum_{i=1}^m \cos\theta_i \ket{i}\bra{i}$ and $S_m \equiv \sum_{i=1}^m \sin\theta_i \ket{i}\bra{i}$}. The direct sum $\mathcal{S}_{2m}\oplus\mathds{1}_{n-m}$ can also be written in the form of a block diagonal matrix
\begin{align}
\begin{pmatrix}
\mathcal{S}_{2m} & 0\\
0& \mathds{1}_{n-m}
\end{pmatrix}\,.
\end{align}
 In this work we will only consider cases where $m=n$, so that 
\begin{equation}
U = \begin{pmatrix}
L & 0\\
0 & L'
\end{pmatrix} \begin{pmatrix}
C & S\\ -S & C
\end{pmatrix} \begin{pmatrix}
R^\dagger & 0\\
0 & R'^{\dagger}
\end{pmatrix}\,,\label{eq:CS_1}
\end{equation}
where, for convenience, we drop the indices which indicate dimension. The matrix $\mathcal{S}$ can be further decomposed, see Appendix (App.)~\ref{Ap:1},  as 
\begin{equation}
\mathcal{S} = \begin{pmatrix}
C & S\\ -S & C
\end{pmatrix}= \Big(P_{\frac{\pi}{4}}^\dagger H\otimes\mathds{1}\Big)
 \begin{pmatrix} \Theta & 0\\ 0 & \Theta^\dagger \end{pmatrix}
\Big(H P_{\frac{\pi}{4}}\otimes\mathds{1}\Big),
\label{eq:bs_internal_decomp}
\end{equation}
where $P_{\phi} =\exp(i\phi) \ket{0}\bra{0}+\exp(-i\phi)\ket{1}\bra{1}$ may represent a phase shift and  {$H= \ket{0}(\bra{0}+\bra{1})/\sqrt{2} + \ket{1}(\bra{0}-\bra{1})/\sqrt{2} $} is the Hadamard transformation -- both acting on a two-level system.  In addition, $\Theta=C+iS$ can be implemented as a state-dependent phase shift of an $n$-level system. 

\begin{figure}
{
\begin{tikzpicture}
\draw[dashed] (0,0.5) -- (0.5,0.5); 
\draw (0.75,0.5) node {$R'^\dagger$};
\draw (0.5,0.25)rectangle(1,0.75);
\draw[dashed] (1,0.5) -- (1.5,0.5);

\draw (0,1.5) -- (0.5,1.5);
\draw (0.75,1.5) node {$R^\dagger$};
\draw (0.5,1.25)rectangle(1,1.75);
\draw (1,1.5) -- (1.5,1.5);

\draw[dashed] (1.5,0.5) -- (2,1);
\draw (2,1) -- (2.5,1.5);
\draw[very thick] (1.75,1) -- (2.25,1);
\draw (1.5,1.5) -- (2,1);
\draw[dashed] (2,1) -- (2.5,0.5);

\draw[dashed] (2.5,0.5) -- (3,0.5);
\draw (3.25,0.5) node {$\Theta^\dagger$};
\draw (3,0.25)rectangle(3.5,0.75);
\draw[dashed] (3.5,0.5) -- (4,0.5);

\draw (2.5,1.5) -- (3,1.5);
\draw (3.25,1.5) node {$\Theta$}; 
\draw (3,1.25)rectangle(3.5,1.75);
\draw (3.5,1.5) -- (4,1.5);

\draw[dashed] (4,0.5) -- (4.5,1);
\draw (4.5,1) -- (5,1.5);
\draw[very thick] (4.25,1) -- (4.75,1);
\draw (4,1.5) -- (4.5,1);
\draw[dashed] (4.5,1) -- (5,0.5);

\draw[dashed] (5,0.5) -- (5.5,0.5); 
\draw (5.75,0.5) node {$L'$};
\draw (5.5,0.25)rectangle(6,0.75);
\draw[dashed] (6,0.5) -- (6.5,0.5);

\draw (5,1.5) -- (5.5,1.5);
\draw (5.75,1.5) node {$L$};
\draw (5.5,1.25)rectangle(6,1.75);
\draw (6,1.5) -- (6.5,1.5);
\end{tikzpicture}
}

\caption{\textbf{Operational meaning of the CS decomposition.} The solid (dashed) line represents the basis state $\ket{0}$ ($\ket{1}$) of the controller.  Each box on the solid (dashed) line depicts an operation acting on the system, given that the controller is in the state $\ket{0}$ ($\ket{1}$). Initially the system is subjected to unitary evolution $R^\dagger$ ($R'^\dagger$) if the controller is in state $\ket{0}$ ($\ket{1}$). Thereafter the controller undergoes a Hadamard transformation (thick horizontal line) followed by a sequence of {conditional} unitaries acting on the system with a second Hadmard transformation of the controller in between.
}
\label{fig:fig1}
\end{figure}
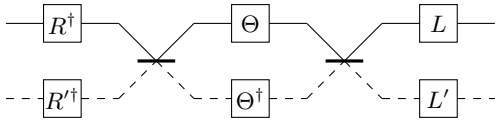

Any unitary operator acting on a $2n$-dimensional Hilbert space can therefore be written 
\begin{equation}\label{eq:CS_U}
U = \begin{pmatrix}
 L & 0\\
0 & iL'
\end{pmatrix} 
H\otimes\mathds{1}
 \begin{pmatrix}
\Theta & 0\\
0 & \Theta^\dagger
\end{pmatrix}
H \otimes\mathds{1}
\begin{pmatrix}
 R^\dagger & 0\\
0 & -i R'^{\dagger}
\end{pmatrix}\,.
\end{equation}
The CS decomposition thus points to the physical realisation of a ($2n \times 2n$) unitary operator in terms of the evolution of a closed system formed by two subsystems, i.e.,  a two-level system coupling to an $n$-level system. This agrees with a sequence of evolutions of a system of $n$ spatial modes of light depending on two paths (as e.g.\ realised in a Mach-Zehnder) or depending on its polarisation.  Hence, the CS decomposition has an operational meaning for a composite system, which is represented in Fig.~\ref{fig:fig1}. 

\section{Optical implementation of coherent feedback}\label{Sec:2-out}
Based on a realisation method for optical qubit channels \cite{SandeepBox21}, we provide a recipe for the design of the optical implementation of coherent feedback given the corresponding Kraus operators from Sec.~\ref{Sec:SFP}.    
The construction of the latter is discussed in  \cite{ konrad2020robust,Uys2018SFP}. We note that the corresponding control channel can also be realised by measurement, constituting measurement-based feedback control \cite{Uys2018SFP}. Therefore what follows is also a recipe to translate measurement-based feedback control into coherent feedback control.  
 
The discussion is limited to the control of two-level systems (qubits) by qubit controllers, corresponding to control channels with a pair of Kraus operators. However, an analogous scheme can be readily  constructed for $d$-level systems controlled by $n$-level ancilla systems, i.e.,\ control channels involving $n$ Kraus operators~\cite{SandeepBox21}. 

A simple deterministic preparation channel for qubits that reaches target state $\ket{T}$ in a single shot, is given by  $K_0=\ket{T}\bra{T}$ and $K_1=\ket{T}\bra{T^\perp}$ with $\langle T^\perp\ket{T}=0$.  In general, let $K_0$ and $K_1$ be Kraus operators which meet conditions~\eqref{eq:fp_contition} and \eqref{eq:span_contition}, from which we can construct a unitary of the form
\begin{equation}
U=\begin{pmatrix}
K_0 & A\\
K_1 & B
\end{pmatrix},
\label{eq:unitary_two_outcome}
\end{equation}
where the matrices $A\equiv \braket{0|U|1}$ and $B\equiv \braket{1|U|1}$ are appropriately chosen such that $U$ is unitary. In addition, Eq.~\eqref{Eq:K_U} prescribes $K_0\equiv \braket{0|U|0}$ and $K_1\equiv \braket{1|U|0}$. This is in agreement with the form of $U$ in Eq.~\eqref{eq:unitary_two_outcome}, which acts on the Hilbert space $\mathcal{H}_c\otimes \mathcal{H}_s$, where $c$ and $s$ stands for controller and system, respectively.

The CS decomposition of $U$ is given by Eq.~\eqref{eq:CS_1}, by comparing this to Eq.~\eqref{eq:unitary_two_outcome} we can write the Kraus operators as
\begin{equation}
\begin{aligned}
K_0 &= L C R^\dagger \\
K_1 &= i L'' S R^\dagger\,,
\end{aligned}
\label{eq:svd_two_outcome}
\end{equation}
where $L''=iL'$. This is simply a singular value decomposition which exists for all Kraus operators $K_i$. Unitary $R'$ can be freely chosen and this can be exploited in order to simplify the optical implementation of $U$. For this purpose, we choose $R' = -iR$. This converts the conditional unitary $R^\dagger \oplus -iR'^\dagger$ to the local unitary $\mathds{1}\otimes R^\dagger$. 

Our aim is now to design an optical set-up to implement the unitary operator
\begin{equation}\label{eq:U_4}
U = \begin{pmatrix}
L & 0\\
0 & L''
\end{pmatrix} 
H\otimes\mathds{1}
 \begin{pmatrix}
\Theta & 0\\
0 & \Theta^\dagger
\end{pmatrix}
H \otimes  R^{\dagger}
\end{equation}
where $L$, $L''$, $R^\dagger$ and $\Theta = C +i S$ are determined by the singular value decomposition of the Kraus operators $K_0$ and $K_1$, given in Eq.~\eqref{eq:svd_two_outcome}. 

Thus far the system and controller degree of freedom have not been specified. Below we discuss two specific examples of controller and system, namely path and polarisation, followed by the polarisation and orbital angular momentum.

\subsection{Path and polarisation}\label{Sec:Pol_real}
\begin{figure}
\includegraphics[width=0.75\columnwidth]{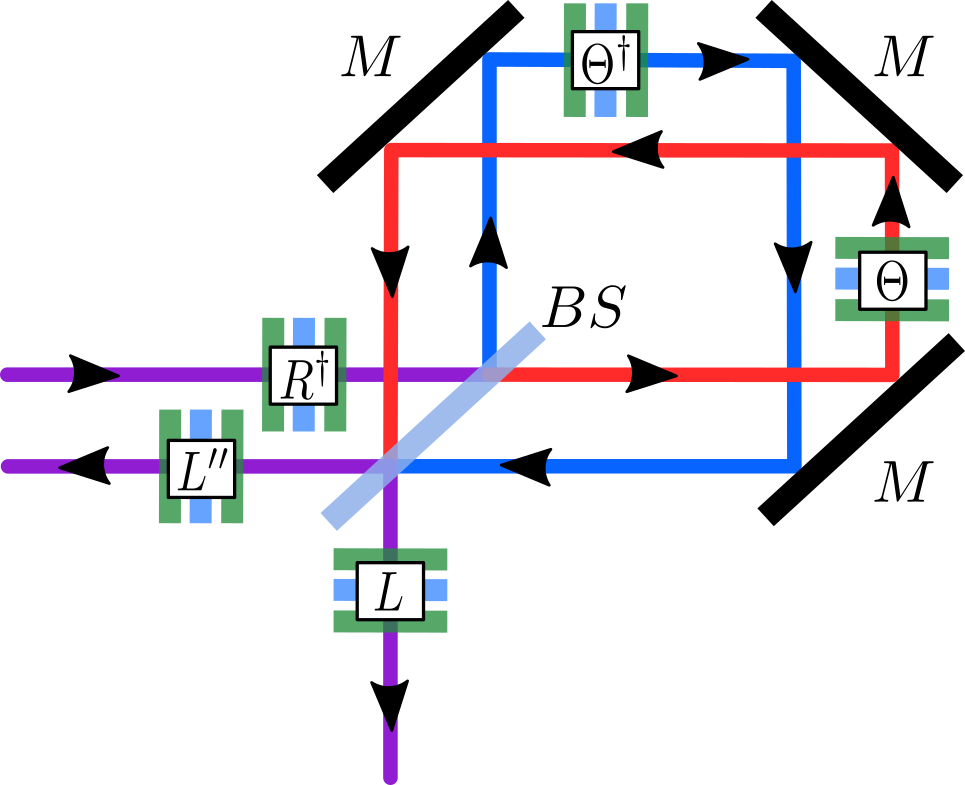}
\caption{\textbf{Optical realisation of coherent feedback control for path and polarisation.} Light entering along a single path (purple) passes through a sequence of quarter-wave plates (green) and half-wave plates (blue)  that implements $R^\dagger$ on the polarisation degree of freedom. A beams-splitter ($BS$) separates the beam into two paths (blue and red). Mirrors are labelled $M$. The light beam exits the interferometer along two paths (purple). } \label{fig:fig_3}
\end{figure}

If we consider the paths of a photon as the controller and its polarisation as the system degree of freedom, then the CS decomposition of $U$, Eq.~(\ref{eq:U_4}),  can be implemented using balanced beam-splitters and local unitary operations on the polarisation. The implementation of the latter require just one half-wave plate and two quarter-wave plates mounted coaxially~\cite{simon_gadget1,simon_gadget2}.  
The decomposition of $U$ represents a generalized Mach-Zehnder interferometer and can be made robust against noise associated with the vibration of optical elements by employing a Sagnac interferometer, shown in Fig.~\ref{fig:fig_3}.

In order to determine the configuration of half- and quarter-wave plates which realises $L$, for example, we first expressed $L$ in terms of Euler angles $(\xi,\eta,\zeta)$,
\begin{align}
L(\xi,\eta,\zeta) &= e^{-i\frac{1}{2}\xi\sigma_y}e^{i\frac{1}{2}\eta\sigma_z}e^{-i\frac{1}{2}\zeta\sigma_y}\,.
\end{align}
The relationship between the Euler angles and the rotation angles of a  half-wave plate ($H_{\phi}$) and two quarter-wave plates ($Q_\phi$) is  given by~\cite{simon_gadget2}:
\begin{align}
L(\xi,\eta,\zeta)& = Q_{\frac{\pi}{4}+\frac{\xi}{2}}  H_{-\frac{\pi}{4}+\frac{\xi+\eta-\zeta}{4}}Q_{\frac{\pi-\zeta}{4} }\,.\label{eq:QQH}
\end{align}  
It is possible to reorder the optical elements by exploiting the fact that
\begin{align}
Q_\phi H_{\phi'} = H_{\phi'}Q_{2\phi'-\phi}\,.
\end{align}
The same decomposition procedure can be performed for $L''$, $\Theta$, $R$ and $R'$.

\subsection{Polarisation and orbital angular momentum}\label{Sec:OAM_real}
In this section we present a linear optical scheme to implement the coherent feedback control of a two-dimensional subspace of the OAM degree of freedom of light. Here the polarisation degree of freedom is considered as controller, hence, the scheme can be made non-interferometric. In order to determine the optical implementation of $U$ for polarisation and OAM we first express $U$ by means of local unitaries and controlled unitaries of the form $C_A := \mathds{1}\oplus A$. Eq.~\eqref{eq:U_4} can be written in terms of controlled unitary operations $C_{ L'' L^\dagger}$ and $C_{(\Theta_2^\dagger)^2}$ as
\begin{align}
U &= \begin{pmatrix}
\mathds{1}  & 0\\
0 & L''L^\dagger
\end{pmatrix} 
H\otimes L\Theta
 \begin{pmatrix}
\mathds{1} & 0\\
0 & (\Theta^\dagger)^2
\end{pmatrix}
H \otimes  R^{\dagger}\nonumber
\\&= C_{ L''L^\dagger} \left(H\otimes L\Theta\right) C_{(\Theta^\dagger)^2}  \left(H \otimes  R^\dagger\right) \,.\label{eq:U_pol_oam}
\end{align}
In order to determine the optical implementation of the controlled unitary $C_{L''L^\dagger}$, it is first diagonalised:
\begin{align}
C_{ L''L^\dagger}  = (\mathds{1}\otimes W) (\mathds{1}\oplus \Phi ) (\mathds{1}\otimes W^\dagger)\,,\label{eq:LL_diag}
\end{align}
where $W\Phi W^\dagger =  L''L^\dagger$ and $\Phi = \sum_j \exp(i\phi_j)\ket{j}\bra{j}$, so that 
\begin{equation}
U = \left(\mathds{1} \otimes  W\right) C_{\Phi} \left(H\otimes W^\dagger L\Theta \right) C_{(\Theta^\dagger)^2}  \left(H \otimes  R^\dagger\right) \,. \label{eq:oam_pol_u}
\end{equation}
The controlled unitary operations $C_{\Phi}$ and $C_{(\Theta^\dagger)^2}$ have polarisation as control bit, where $\ket{0}\equiv \ket{V}$ (vertically polarised light) and $\ket{1}\equiv \ket{H}$ (horizontally polarised light), and as target bit the OAM subspace spanned by $\{\ket{-\ell}\equiv \ket{0},\ket{+\ell}\equiv \ket{1}\}$. Such controlled unitaries can be implemented using a linear optical device named polarisation selective Dove prism (PSDP)~\cite{yasir2021polarization}. 

\begin{figure}[t]
\begin{subfigure}{0.5\textwidth}
\includegraphics[width=0.6\textwidth]{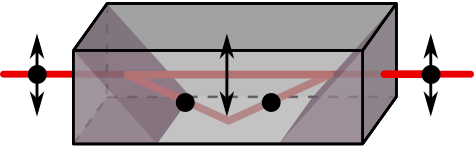}
\caption{\textbf{Modified Dove prism.} Three uniaxial crystals are sandwiched together to form a cuboid. All three components are constructed from the same uniaxial crystal, however their optic axes are aligned in different directions~\cite{yasir2021polarization}. Vertically polarised light (\,\rotatebox[origin=c]{90}{$\leftrightarrow$}\,) passes through the device, while horizontally polarised light (\,$\bullet$\,) is reflected inside the trapezoid (grey). }\label{fig:psdp} 
\end{subfigure}
\begin{subfigure}{0.5\textwidth}
\includegraphics[width=0.6\textwidth]{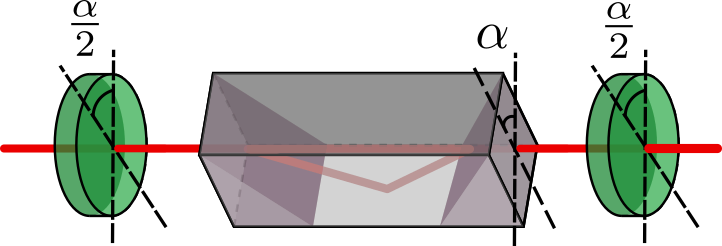}
\caption{\textbf{Modified Dove prism between two quarter-wave plates forms the PSDP device.} The rotated PSDP which implements the controlled operation in Eq.~\eqref{Eq:PSDP} consists of a modified Dove prism rotated by an angle $\alpha$, between two quarter-wave plates that are each rotated by $\alpha/2$.}\label{fig:psdp_rot}
\end{subfigure}
\caption{Diagram showing the components of the polarisation selective Dove prism (PSDP).}\label{figthree}
\end{figure}
A PSDP consists of a modified Dove Prism (DP) mounted between two half-wave plates, cp.\ Fig.\ \ref{figthree}. A rotated PSDP realises the following controlled operation
\begin{align}
\mathcal{C}^{PSDP}_{l,\alpha} := \begin{pmatrix}
\mathds{1} & 0 \\ 0 & \sigma_x P_{2\ell\alpha}^\dagger
\end{pmatrix} = \begin{pmatrix}
\mathds{1} & 0 \\ 0 & P_{2\ell\alpha} \sigma_x 
\end{pmatrix} \,, \label{Eq:PSDP} 
\end{align}
where $\alpha$ is the angle of rotation and $P_{2\ell\alpha}^\dagger$ is the previously defined phase shift operator. The rotated PSDP device is configured by rotating the modified DP by an angle $\alpha$ and the half-wave plates by $\alpha/2$. 

It is therefore possible to implement the controlled unitaries $C_\Phi$ and $C_{(\Theta^\dagger)^2}$ using a local phase shift on polarisation and two PSDPs. This is further outlined in App.~\ref{Ap:2}. If $L'L^\dagger$ can be written in the form $\exp(i2\phi)\sigma_x P^\dagger_{\theta}$, then a phase shifter implementing $P^\dagger_{\phi}$ on polarisation and a single rotated PSDP can be used to implement $C_{L'L^\dagger}$ in Eq.~\eqref{eq:U_pol_oam}, without the need to diagonalise it.

The Hadamard operation on polarisation can be performed using a half-wave plate rotated by an angle $\pi/8$. The final requirement for the optical realisation of $U$ is local unitary operations on the OAM degree of freedom. It is not currently known how to implement arbitrary unitary operations on OAM. However, cylindrical-lens mode converters~\cite{beijersbergen1993astigmatic} can be used to perform unitary operations on the $\{\ket{\ell=-1},\ket{\ell=+1}\}$ subspace of OAM. The so called $\pi$- and $\pi/2$-converters are analogous to the half-wave and quarter-wave plates, respectively~\cite{padgett1999poincare} and can therefore be used to construct arbitrary unitary operations on this subspace of OAM. In addition, a rotated Dove prism can be used to implement the $P_{2\ell\alpha}\sigma_x$ operation on the two-level OAM subspace associated with $\ell$, where $\alpha$ is the angle of rotation of the Dove prism.


\section{Iterative application of control channels}\label{Sec:Reuse}
In Sec.~\ref{Sec:2-out} we showed how to construct coherent feedback  control schemes from control channels which satisfy the conditions for convergence to a target state given in Eqs.~\eqref{eq:fp_contition} and \eqref{eq:span_contition}. Such schemes may require repetition for various reasons. Some control schemes work weakly, meaning convergence to the target state occurs over multiple iterations. Examples of weak schemes are given in Sec.~\ref{Sec:weakswap} and Sec.~\ref{Sec:Example_2}. In addition, repeated coherent feedback control protects the system against noise~\cite{konrad2020robust}. Here we discuss two methods to achieve repeated application of coherent feedback control.

If the system starts in the pure state $\ket{\psi_0}$,  the initial state of system and controller is given by $\ket{\Psi_0}:=\ket{0}\ket{\psi_0}$. The output state $\ket{\Psi_1}:= U \ket{\Psi_0}$ of a coherent feedback scheme then reads
\begin{equation}
    \begin{aligned}
    \ket{\Psi_1} & = \sum_i \ket{i}\bra{i}U \ket{0}\ket{\psi_0} \\
    &= \ket{0} K_0 \ket{\psi_0}+ \ket{1} K_1 \ket{\psi_0}\,, 
    \end{aligned}\label{bla0}
\end{equation}
where the Kraus operators $K_0$ and $K_1$ obey the conditions ~\eqref{eq:fp_contition} and \eqref{eq:span_contition} to drive the system into a target state. 
In order to iterate the control channel, the controller must be reset to its initial state $\ket{0}$.

This could be accomplished by submitting each of the two components in Eq.\ (\ref{bla0}) into an additional control device adjusting the state of the controller where necessary.  However,  this would require $2^{N}-1$ devices for $N$ iterations.  We discuss in the following two ways to circumvent this exponential growth of machinery. 
The first method involves the resetting of the controller by filtering. This results in losses that we propose to compensate by parametric amplification. The second method involves the transfer of the controller state to an additional degree of freedom. 

\subsection{Resetting of controller by filtering}\label{reset}
One way to reset the controller is to project it onto the state $\ket{+}\equiv\tfrac{1}{\sqrt{2}}(\ket{0}+\ket{1})$ to create a superposition of the Kraus operators:
\begin{equation}
   \ket{\Psi_1}\xrightarrow{\ket{+}\bra{+}} \ket{\Psi_1'} = \frac{1}{\sqrt{2}} \ket{+} (K_0 \ket{\psi}+K_1 \ket{\psi}) .
   \label{project}
\end{equation}
A subsequent rotation restores the initial controller state $\ket{0}$,
\begin{equation}
    \ket{\Psi_1'} \xrightarrow{} \ket{\Psi_1''}= \frac{1}{\sqrt{2}}\ket{0}(K_0 \ket{\psi}+ K_1\ket{\psi}).
\end{equation}
After the projection (\ref{project}) the information about the controller states is deleted, and the state change no longer corresponds to a trace-preserving channel that satisfies the conditions for convergence given in Eqs.~\eqref{eq:fp_contition} and~\eqref{eq:span_contition}. However, we show that for the specific examples of coherent feedback schemes discussed in Sec.~\ref{Sec:Examples}, the system will still converge to the target state (see App.~\ref{Ap:fid}).
The state of the composite system after the second round of coherent control is given by $\ket{\Psi_2}:=U\ket{\Psi_1''}$. The projection \eqref{project} will, in general, lead to photon losses. In order to compensate for these we consider parametric amplification.

The preparation of light modes or polarisation by coherent feedback control can be applied to coherent (classical) states of light. In this case, amplification is an adequate means to compensate for losses. Parametric amplification is available for both spatial light modes and polarisation as system degree of freedom. This technique uses a non-linear crystal and a pump beam to amplify an input signal beam. The intensity gain factor of the parametric amplifier can be appropriately adjusted in order to restore the original input beam intensity.

The parametric intensity gain is given by~\cite{opampManzoni},
\begin{equation}
    G(L)= 1 + \left(\frac{\Gamma}{g} \sinh(g L) \right)^2. 
\end{equation}
Here $L$ is the length of the  crystal (interaction length), and  $g$ and $\Gamma$ are generalised wave numbers that depend on the parameters of the non-linear process, i.e., the pump beam intensity, the phase matching, the angular frequencies, the wave numbers in the medium as well as the non-linear susceptibility (see App.~\ref{Ap:4}). 

Resetting by means of filtering is available for different controller degrees of freedom. For the path degree of freedom, the light from the two output ports of the interferometer shown in Fig.~\ref{fig:fig_3} can be combined according to Eq.~\eqref{project} using a balanced beam-splitter. However, one path will still have to be discarded and this intensity loss could be compensated by amplification. For polarisation, the intensity losses are due to the use of  a linear polariser (filter) which performs the projection onto $(\ket{V}+\ket{H})/\sqrt{2}$ (Eq.~\eqref{project}). A polarisation rotator can then be employed to reset the initial polarisation $\ket{V}$.

We are not limited to the use of classical light, as amplification still works using low numbers of photons, but with finite success probabilities. In this regime, the output photons would only reach a limited fidelity, in agreement with the no-cloning theorem. It has been shown in \cite{quantclon} that stimulated emission, and thus parametric amplification, is capable of producing quantum clones with near optimal fidelity. For an optimal universal $N \rightarrow M$ cloner, the optimal fidelity is given by \cite{qclonfid},
 \begin{equation}\label{Eq:opfid}
     F= \frac{NM +N +M}{M (N+2)}.
 \end{equation}
 For the case of $N,M\rightarrow\infty$ in Eq.~\eqref{Eq:opfid}, the classical fidelity of $1$ is recovered.
 
 Possibly, the noise present in the amplification process on the single photon level is reduced with each repetition of the control channel. This may lead to a high fidelity asymptotically, but with finite success probability. However, we do not know whether the coherent feedback method introduced here works for non-classical states of light with more than one photon. This is subject of further investigation.

\subsection{Storage in an additional degree of freedom}\label{Sec:discard}    
 Here we make use of an additional degree of freedom in order to reset the controller. Taking into account an additional ancilla ``$a$" in initial state $\ket{0}_a$, the composite state $\ket{\Psi_1}:= (U\ket{0}_c\ket{\psi_0}_s)\ket{0}_a$ after the system interacted with the controller reads 
\begin{equation}
  \ket{\Psi_1 }= (\ket{0}  K_0 \ket{\psi_0} + \ket{1} K_1 \ket{\psi_0}) \ket{0}_a\,
\end{equation}
where $\ket{\psi_0}$ is the initial state of the system. In a first step, we mark the controller basis states using basis states of the ancilla $a$, e.g., by a C-NOT operation acting on ``$a$" conditioned on the controller,
\begin{equation}
\ket{\Psi_1 } \rightarrow \ket{\Psi_1'} = \ket{0} K_0 \ket{\psi}\ket{0}_a + \ket{1} K_1 \ket{\psi}\ket{1}_a.\label{Eq:marking}
\end{equation}

In this case, the initial controller state can be restored by a C-NOT conditioned on the state of subsystem ``$a$":
\begin{equation}
  \ket{\Psi_1'} \rightarrow \ket{\Psi_1''}= \ket{0} \left(K_0 \ket{\psi} \ket{0}_a +  K_1\ket{\psi}\ket{1}_a\right).\label{Eq:reset}
\end{equation}
In order to restore the initial controller state $\ket{0}$ unitarily, the information about the unknown state of the controller after the application of the control channel must be stored in the ancillary degree of freedom. When this resetting process is extended to $N$ iterations of the control channel, the dimension of the ancilla is given by $2^N$.
Below we discuss this mechanism at the example of the time-bins and OAM as ancillas.    

\subsubsection{Time-bins as additional degree of freedom}\label{Sec:timebin}
\begin{figure}
\centering
\def\svgwidth{\columnwidth}
\begingroup%
  \makeatletter%
  \providecommand\color[2][]{%
    \errmessage{(Inkscape) Color is used for the text in Inkscape, but the package 'color.sty' is not loaded}%
    \renewcommand\color[2][]{}%
  }%
  \providecommand\transparent[1]{%
    \errmessage{(Inkscape) Transparency is used (non-zero) for the text in Inkscape, but the package 'transparent.sty' is not loaded}%
    \renewcommand\transparent[1]{}%
  }%
  \providecommand\rotatebox[2]{#2}%
  \newcommand*\fsize{\dimexpr\f@size pt\relax}%
  \newcommand*\lineheight[1]{\fontsize{\fsize}{#1\fsize}\selectfont}%
  \ifx\svgwidth\undefined%
    \setlength{\unitlength}{406.30851216bp}%
    \ifx\svgscale\undefined%
      \relax%
    \else%
      \setlength{\unitlength}{\unitlength * \real{\svgscale}}%
    \fi%
  \else%
    \setlength{\unitlength}{\svgwidth}%
  \fi%
  \global\let\svgwidth\undefined%
  \global\let\svgscale\undefined%
  \makeatother%
  \begin{picture}(1,0.57544573)%
    \lineheight{1}%
    \setlength\tabcolsep{0pt}%
    \put(0,0){\includegraphics[width=\unitlength,page=1]{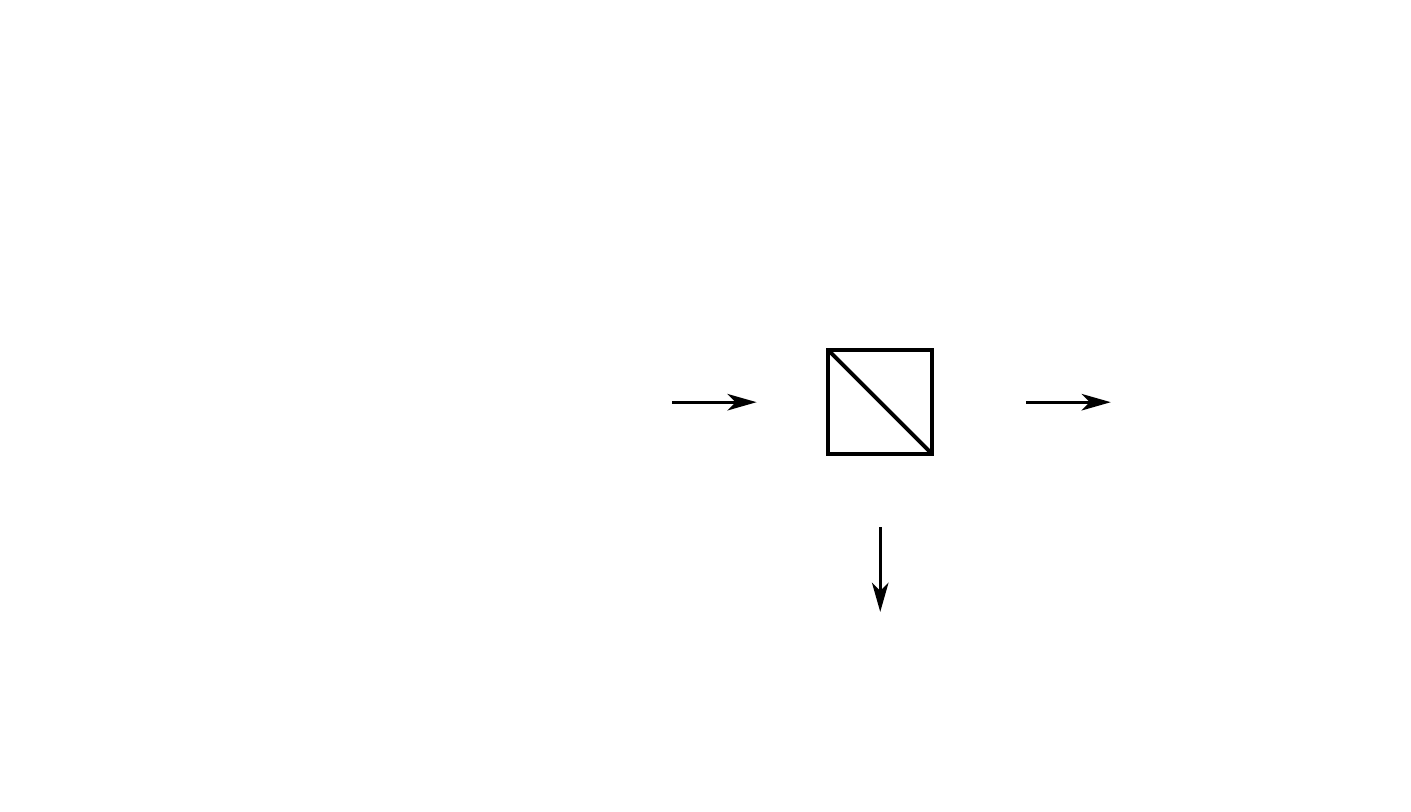}}%
    \put(0.7161792,0.53957766){\color[rgb]{0,0,0}\makebox(0,0)[lt]{\lineheight{1.25}\smash{\begin{tabular}[t]{l}EOM1\end{tabular}}}}%
    \put(0.14179634,0.3205866){\color[rgb]{0,0,0}\makebox(0,0)[lt]{\lineheight{1.25}\smash{\begin{tabular}[t]{l}EOM2\end{tabular}}}}%
    \put(0.63623984,0.22022151){\color[rgb]{0,0,0}\makebox(0,0)[lt]{\lineheight{1.25}\smash{\begin{tabular}[t]{l}PBS\end{tabular}}}}%
    \put(0.29717845,0.41611744){\color[rgb]{0,0,0}\makebox(0,0)[lt]{\lineheight{1.25}\smash{\begin{tabular}[t]{l}coherent\\feedback\end{tabular}}}}%
    \put(0.70552712,0.32924576){\color[rgb]{0,0,0}\makebox(0,0)[lt]{\lineheight{1.25}\smash{\begin{tabular}[t]{l}delay loop\end{tabular}}}}%
    \put(0,0){\includegraphics[width=\unitlength,page=2]{figure4.pdf}}%
  \end{picture}%
\endgroup%

\caption{Setup to create disjoint time-bins depending on the order of Kraus operators in $N$ iterations of the channel which is realised by coupling polarisation (controller) to OAM (system). The setup uses electro-optic modulators (EOM) that change the polarisation of a pulse when switched on, and a polarising beam-splitter (PBS) that reflects (transmits) vertically (horizontally) polarised light~\cite{EOMswitch}. EOM1 keeps the horizontally polarised component in the delay loop for the required number of rounds in each iteration. EOM2 ensures that the polarisation of each time-bin is reset to the initial state before re-entering the channel.}
\label{fig:time}
\end{figure}

We consider coherent feedback control with polarisation and OAM of a light pulse as the controller and the system, respectively. The ancilla (time-bins) is in initial state $\ket{t=0}$, where $t$ represents the time delay between light pulses. Distinct time-bins can be created by varying the path lengths that the two different polarisation components experience by means of a delay loop. This marks each polarisation state with a time-bin state, analogous to Eq.~\eqref{Eq:marking}. The initial polarisation can then be restored in each pulse before it re-enters the coherent feedback scheme for the next iteration (Eq.~\eqref{Eq:reset}). Given that the output state of the pulse after the first application of the coherent feedback is $\ket{\Psi_{1}}:=(U\ket{V}_c\ket{\psi_0}_s)\ket{0}_a$, then the total transformation described can be written as
\begin{equation}\label{timetrans}
  \begin{aligned}
  &\ket{\Psi_{1}}=\ket{V}  K_0 \ket{\psi_0}\ket{0}_a+ \ket{H} K_1 \ket{\psi_0}\ket{0}_a\\
  & \longrightarrow \ket{\Psi_1''}= \ket{V} (K_0 \ket{\psi_0}\ket{0}_a+ K_1 \ket{\psi_0}\ket{ \tau}_a)\,, 
  \end{aligned}
\end{equation}
where $\tau$ is a delay period.
A proposed method to iteratively implement Eq.~\eqref{timetrans} is shown in Fig.~\ref{fig:time}.

The delay time $\tau$ must change in each iteration, in such a way that the pulses do not overlap. This allows for the selective change of the polarisation in order to restore the initial controller state for each time-bin. One way to create distinct pulses is to double the number of round trips in the delay loop from one iteration to the next. This is demonstrated in App.~\ref{unique}.

 During the second iteration the two distinct pulses in $\ket{\Psi_1''}$ are taken (at different times) as inputs in the coherent feedback scheme so that the state $\ket{\Psi_{2}}:= U\ket{\Psi_1''}$ is given by 
\begin{equation}
    \begin{aligned}
    \ket{\Psi_{2}} =& \left(\ket{V} K_0^2\ket{\psi} + \ket{H} K_1 K_0\ket{\psi}\right)\ket{0}_a\\
    & + \left(\ket{V} K_0 K_1 \ket{\psi}+ \ket{H} K_1^2\ket{\psi}\right)\ket{\tau}_a. 
    \end{aligned}
\end{equation}
The polarisation states are marked by the conditional operation on the controller and ancilla ``$a$", 
\begin{align}
\ket{V}\bra{V}\otimes\mathds{1}+\ket{H}\bra{H}\otimes\sum_{n=0}^{2^{N-1}-1}\ket{n\tau+2^{N-1}\tau}\bra{n\tau}\,,
\end{align} 
where $N$ is the number of applications of the coherent feedback control which have already occurred, and therefore corresponds to the index of the composite state. Here $N=2$ so that
\begin{equation}
    \begin{aligned}
    \ket{\Psi_2'}=& \ket{V}K_0^2\ket{\psi}\ket{0}_a  + \ket{H} K_1 K_0\ket{\psi}\ket{2\tau}_a\\
    & + \ket{V} K_0 K_1 \ket{\psi}\ket{\tau}_a+ \ket{H} K_1^2\ket{\psi}\ket{3\tau}_a. 
    \end{aligned}
\end{equation}

It is clear that the leading pulses in the first half of the pulse train are in the desired polarisation state. However, the second half of all pulses in the pulse train require resetting. This can be achieved by the conditional operation on the controller and ancilla ``$a$", 
\begin{align}
\mathds{1}\otimes \sum_{n=0}^{2^{N-1}-1} \ket{n\tau}\bra{n\tau} + \sigma_x\otimes \sum_{n=2^{N-1}}^{2^N-1}\ket{n\tau}\bra{n\tau}\,,
\end{align}
so that
\begin{equation}
    \begin{aligned}
    \ket{\Psi_2''}= \ket{V} &\left( K_0^2\ket{\psi}\ket{0}_a + K_1 K_0\ket{\psi} \ket{\tau}_a \right.\\
    & \quad \left.+ K_0 K_1 \ket{\psi}\ket{2\tau}_a + K_1^2\ket{\psi}\ket{3\tau}_a\right). 
    \end{aligned}
\end{equation}

After a sufficiently high number $N$ of applications of the control channel, the system converges to the target state $\ket{T}$ \cite{konrad2020robust}. Therefore, after $N$ iterations of coherent control and controller reset the composite system is in a product state of the form  
\begin{equation}
    \ket{\Psi_N''}=\ket{V} \ket{T}\left(\sum_{m=0}^{2^N-1} \alpha_m \ket{m\tau}_a\right).
\end{equation}
The amplitudes of the $2^N$ pulses are given by  $\alpha_m= \bra{T} \mathcal{K}_m\ket{\psi}$ as follows from writing the final state in terms of the initial state,  $\ket{\Psi_N''}= \ket{V} \left(\sum_m  \mathcal{K}_m\ket{\psi_0}\ket{m\tau}_a\right)$. Here $\mathcal{K}_m$ refers to the $m$th permutation of the product $\Pi_{j=1}^N K_{i}^{(j)}$ of the two Kraus operators, with $i=0,1$.

Although the system degree of freedom of each pulse is in the target state, the pulses have low amplitudes. In measurements on the system, collecting the accumulated signal over a large time period compensates the low amplitudes. 

\subsubsection{OAM as additional degree of freedom} \label{Sec:oam_mul}
We consider coherent feedback control with path as the controller, polarisation as the system, and employ OAM as an additional ancillary degree of freedom. The input state has an even OAM mode so that the composite system is $\ket{0}_c\ket{\psi_0}_s\ket{\ell=0}_a$. Here $\ket{0}_c$ corresponds to the initial state of the path degree of freedom, which is the upper path, and $\ket{\psi_0}_s$ is the initial state of the system. 

After the coherent feedback, one unit of OAM is added to the light in the lower path (with $\ket{1}_c$)  by use of a spiral phase plate. This is done in order to mark the controller, analogous to Eq.~\eqref{Eq:marking}.  The controller can then be reset using an inverse even-odd OAM mode sorter,  so that the total transformation is given by
\begin{equation}\label{eq:oamtrans}
    \begin{aligned}
    &\ket{\Psi_{1}}=(\ket{0}K_0 \ket{\psi_0} + \ket{1}K_1 \ket{\psi_0})\ket{\ell=0}_a \\
    & \longrightarrow\ket{\Psi_1''}= \ket{0}(K_0 \ket{\psi_0}\ket{\ell=0}_a + K_1 \ket{\psi_0}\ket{\ell=1}_a). 
    \end{aligned}
\end{equation}

For the next iteration of the coherent feedback control, the OAM values must be doubled as to only obtain even OAM values. To accomplish this we must have a method of multiplying the OAM values of an input beam. In \cite{oam_mult}, multiplication is achieved by using superpositions of circular-sector transformations of the input beam. This method works best for low OAM values, as the conversion efficiencies decrease for increasing OAM values e.g., when doubling the values $\ell=+1$, $\ell=+2$ and $\ell=+3$, conversion efficiencies are $0,97$, $0,93$ and $0,86$, respectively \cite{oam_mult}.

\begin{figure}
\centering
\def\svgwidth{\columnwidth}
\begingroup%
  \makeatletter%
  \providecommand\color[2][]{%
    \errmessage{(Inkscape) Color is used for the text in Inkscape, but the package 'color.sty' is not loaded}%
    \renewcommand\color[2][]{}%
  }%
  \providecommand\transparent[1]{%
    \errmessage{(Inkscape) Transparency is used (non-zero) for the text in Inkscape, but the package 'transparent.sty' is not loaded}%
    \renewcommand\transparent[1]{}%
  }%
  \providecommand\rotatebox[2]{#2}%
  \newcommand*\fsize{\dimexpr\f@size pt\relax}%
  \newcommand*\lineheight[1]{\fontsize{\fsize}{#1\fsize}\selectfont}%
  \ifx\svgwidth\undefined%
    \setlength{\unitlength}{406.1270976bp}%
    \ifx\svgscale\undefined%
      \relax%
    \else%
      \setlength{\unitlength}{\unitlength * \real{\svgscale}}%
    \fi%
  \else%
    \setlength{\unitlength}{\svgwidth}%
  \fi%
  \global\let\svgwidth\undefined%
  \global\let\svgscale\undefined%
  \makeatother%
  \begin{picture}(1,0.42751912)%
    \lineheight{1}%
    \setlength\tabcolsep{0pt}%
    \put(0,0){\includegraphics[width=\unitlength,page=1]{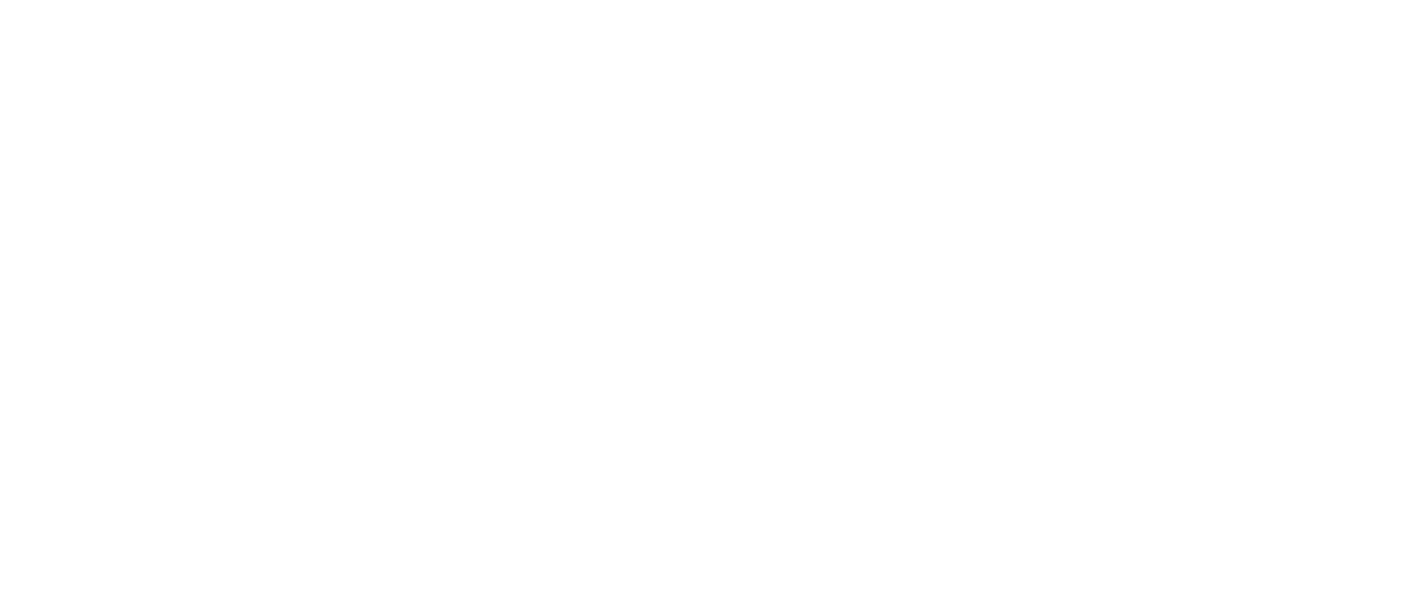}}%
    \put(0.29235272,0.31556606){\color[rgb]{0,0,0}\makebox(0,0)[lt]{\lineheight{1.25}\smash{\begin{tabular}[t]{l}coherent\\feedback\end{tabular}}}}%
    \put(0,0){\includegraphics[width=\unitlength,page=2]{figure5.pdf}}%
    \put(0.64454407,0.06067207){\color[rgb]{0,0,0}\makebox(0,0)[lt]{\lineheight{1.25}\smash{\begin{tabular}[t]{l}S\end{tabular}}}}%
    \put(0.8901194,0.21901899){\color[rgb]{0,0,0}\makebox(0,0)[lt]{\lineheight{1.25}\smash{\begin{tabular}[t]{l}M\end{tabular}}}}%
    \put(0.1688598,0.25865282){\color[rgb]{0,0,0}\makebox(0,0)[lt]{\lineheight{1.25}\smash{\begin{tabular}[t]{l}D\end{tabular}}}}%
  \end{picture}%
\endgroup%

\caption{Setup to reset controller degree of freedom for coherent feedback with polarisation (system) and path (controller) using OAM as additional controller. The marking of the controller states is achieved by S, a spiral phase plate, on the lower path. The initial controller state is restored through the use of M, the inverse OAM mode sorter for even and odd OAM values. Before the next application of the coherent feedback, the OAM values of the light must be doubled with the OAM doubler D, so only even OAM modes are present.}
\label{fig:oammult}
\end{figure}
The setup to iteratively implement the resetting transformation (Eq.~\eqref{eq:oamtrans}) is shown in Fig.~\ref{fig:oammult}. A description of the $N$th iteration of the transformation is now briefly considered. The conditional operation on the controller and ancilla ``$a$" that marks the path states with different OAM values reads 
\begin{equation}
    \sum_{\ell=0}^{2^{N}-2} \left( \ket{0}\bra{0}\otimes\ket{\ell}\bra{\ell}+\ket{1}\bra{1}\otimes\ket{\ell+1}\bra{\ell}\right).
\end{equation}
Since the ingoing OAM modes are all even, the upper path will contain only even OAM modes, while the lower path contains only odd OAM modes.
The initial path state is restored by 
\begin{equation}
    \sum_{l=0}^{2^{N-1}-1} \left(\mathds{1}\otimes\ket{2\ell}\bra{2\ell}+\sigma_x\otimes\ket{2\ell+1}\bra{2\ell+1}\right), 
\end{equation}
which can be implemented by a mode sorter for even and odd OAM values run in reverse.
Before entering the coherent feedback device the OAM value of the light is doubled. 

The system converges to the target state $\ket{T}$  for a sufficiently high number $N$ of iterations as described in the previous subsection. The composite system, after $N$ iterations of  coherent feedback and controller reset but before the final doubling of OAM values, 
is given by
\begin{equation}
    \ket{\Psi_N''}= \ket{0} \ket{T}\left(\sum_{m=0}^{2^{N}-1} \alpha_m \ket{\ell=m}_a\right).
\end{equation}
Each OAM mode has an amplitude $\alpha_m= \bra{T} \mathcal{K}_m\ket{\psi}$. After convergence to the target state, the output yields a scalar beam with the desired polarisation. In this case, it is easy to discard the OAM degree of freedom, as we are only interested in the polarisation.

\section{Examples of coherent feedback schemes }\label{Sec:Examples}
In this section we discuss three coherent feedback schemes. The first is a basic control mechanism which allows a target state to be reached in a single iteration. The second is based on the swap operation and allows any target state to be reached provided it is encoded in the controller degree of freedom.  The third scheme allows for the preparation of an equal superposition of $+\ell$ and $-\ell$ OAM states.

\subsection{Basic control scheme}\label{Sec:Basic}

Let us consider the simple control channel mentioned in Sec.~\ref{Sec:2-out}, with the Kraus operators $K_0= \ket{T}\bra{T}$ and $K_1=\ket{T}\bra{T^\perp}$. This scheme allows the target state to be reached in a single iteration. The singular value decomposition of the Kraus operators reads, $K_0= V \ket{0}\bra{0}V^\dagger$ and $K_1= iV\sigma_y \ket{1}\bra{1}V^\dagger$, where $V = \ket{T}\bra{0} +\ket{T^\perp}\bra{1}$. Hence, according to Eq.~\eqref{eq:U_4}, the CS-decomposition of the unitary that implements the control channel determined by $K_0$ and $K_1$ is given by
\begin{align}
U & = \mathds{1}\otimes V 
\begin{pmatrix}
\mathds{1} & 0\\
0 & \sigma_y
\end{pmatrix} 
H \otimes \mathds{1}
 \begin{pmatrix}
e^{i\frac{\pi}{4}} P_{\frac{\pi}{4}}^\dagger & 0\\
0 & e^{-i\frac{\pi}{4}} P_{\frac{\pi}{4}}
\end{pmatrix}
H \otimes  V^\dagger \,. \label{Eq:simple_example}
\end{align}
If polarisation is the system degree of freedom and the path is the controller, then the above unitary can be implemented using balanced beam-splitters as well as half- and quarter-wave plates as described in Sec.~\ref{Sec:Pol_real}. 
If the system degree of freedom is OAM and the controller is polarisation, then the scheme could be realised using local operations in polarisation and PSPDs as described in Sec.~\ref{Sec:OAM_real}. In the simplest case the target state is given by $\ket{0} \equiv \ket{-\ell}$, so that $V=\mathds{1}$ and (up to a global phase factor)
\begin{align}
U & =  \begin{pmatrix}
\mathds{1} & 0\\
0 & \sigma_x P^\dagger_{\frac{\pi}{2}}
\end{pmatrix} 
P^\dagger_{\frac{\pi}{4}}HP_{\frac{\pi}{4}} \otimes\mathds{1}
 \begin{pmatrix}
\mathds{1} & 0\\
0 & P_{\frac{\pi}{2}}
\end{pmatrix}
H \otimes  P_{\frac{\pi}{4}}^\dagger \,.\label{eq:simlpe_gen}
\end{align}
The local operations $P^\dagger_{\frac{\pi}{4}}HP_{\frac{\pi}{4}}$ and $H$ on polarisation can be decomposed into half- and quarter-wave plates as $Q_{\frac{\pi}{2}}H_{\frac{\pi}{8}}  Q_0$ and $H_{\frac{\pi}{8}}$, respectively. The operations which act on OAM are $\ell$-value dependent. The first controlled unitary in Eq.~\eqref{eq:simlpe_gen} is realised by a PSDP rotated by $\frac{\pi}{4\ell}$, according to Eq.~\eqref{Eq:PSDP}. The second controlled unitary requires two PSDPs mounted coaxially. Finally, the local phase operation $P_{\frac{\pi}{4}}^\dagger$ on OAM can be realised by a two coaxially mounted Dove prisms, one rotated by $\frac{\pi}{8\ell}$.

This example is in essence a mechanism to prepare the system in the state $\ket{0}$. The local unitary $V^\dagger$ can be absorbed into the initial (possibly unknown) system state, while $V$ simply rotates $\ket{0}$ to the target state. This local operation may not be readily available for OAM, as discussed earlier, which limits the applications of this control scheme. In the following examples we explore optical set-ups which allow for more general target states to be reached without the use of inaccessible local OAM operations.

\begin{figure*}[ht]
\begin{subfigure}{0.9\textwidth}
	\includegraphics[width=\textwidth]{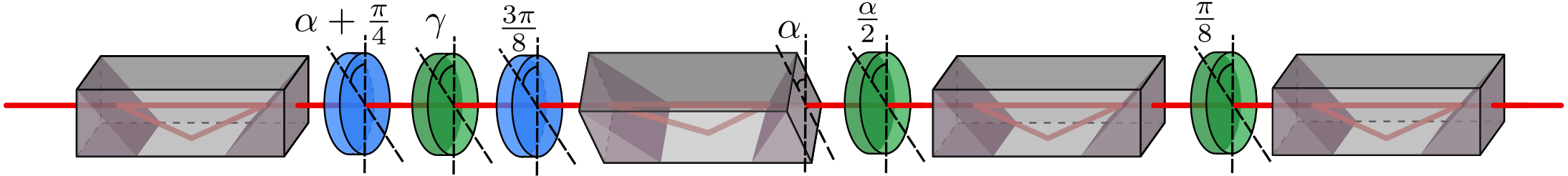}
	\caption{Implementation of the weak swap, where $\gamma = \frac{2\alpha+\lambda}{4}-\frac{\pi}{8}$.}\label{fig:weakswap}
\end{subfigure}
\begin{subfigure}{0.9\textwidth}
	\includegraphics[width=\textwidth]{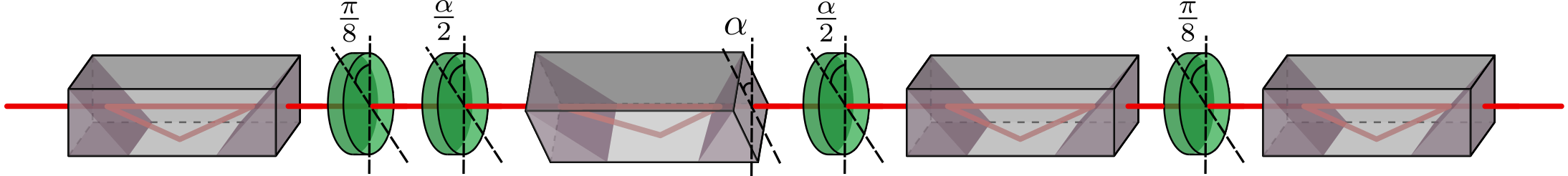}
	\caption{Implementation of the target state dependent control scheme. This optical set-up allows the state $(\ket{-\ell}+\ket{+\ell})/\sqrt{2}$ to be prepared if the polarisation state (controller) is in an equal superposition of vertical and horizontal polarisation.}\label{fig:example2}
\end{subfigure}
\caption{Optical implementation of two coherent feedback control schemes on polarisation (controller) and the subspace of OAM spanned by $\{\ket{-\ell},\ket{+\ell}\}$ (system). Half-wave plates are shown in green, quarter-wave plates in blue and modified Dove prisms are represented by a cuboid. The angle $\alpha=\frac{\lambda}{2\ell}$ depends on the parameter $\lambda$ and on the OAM subspace considered. }\label{figfive}
\end{figure*}

\subsection{Optical implementation of the weak swap}\label{Sec:weakswap}
In~\citep{konrad2020robust} it is shown that the so called weak swap unitary 
\begin{align}
U = \exp\left(-i\lambda S\right)\,,
\end{align}
where $\lambda \in \mathbb{R}$ and $S$ is the swap operator leads to convergence to any target state $\ket{T}$ encoded in the controller system since the Kraus operators $\{K_i = \braket{i|U|T}_c\}_i$ satisfy condition~\eqref{eq:fp_contition} and \eqref{eq:span_contition}. By adding a unitary transformation of the controller state, $U_c\ket{0}_c= \ket{T}_c$, to the weak swap $U\rightarrow U U_c\otimes \mathds{1}$, this case can be reduced to coherent control with initial controller state $\ket{0}_c$. It might be worth noting, that a scheme where the state of the controller determines the target, allows the preparation of an unknown target state that might be the result a quantum computation.  The CS decomposition of the weak swap on $\mathcal{H}_c\otimes \mathcal{H}_s = \mathcal{H}_2\otimes \mathcal{H}_2$ (up to a phase factor) is  given by
\begin{align}
U& = \begin{pmatrix}
  \mathds{1}& 0\\ 0&\sigma_x
\end{pmatrix} 
( H  \otimes \mathds{1} ) 
 \begin{pmatrix} \mathds{1} & 0\\  0& P_{\lambda}^\dagger \end{pmatrix}
(P^\dagger_{\frac{\lambda}{2}} H \otimes  \mathds{1} )
\begin{pmatrix}
  \mathds{1}& 0\\ 0&\sigma_x
\end{pmatrix}\,,\label{Eq:swap_dec}
\end{align}
where $L = P^\dagger_{\frac{\lambda}{2}}$, $  L''=\sigma_xP^\dagger_{\frac{\lambda}{2}}$, $\Theta =  e^{-i\frac{\lambda}{2}}P_{\frac{\lambda}{2}}$, $R =\mathds{1}$ and $R' =\sigma_x $.
The implementation of the weak swap for path (controller) and polarisation (system) can be achieved using beam-splitters, half- and quarter-wave plates, as discussed in Sec.~\ref{Sec:Pol_real}.

The optical implementation of the weak swap for polarisation (controller) and a two-dimensional subspace of OAM spanned by $\{\ket{-\ell},\ket{+\ell}\}$ (system) is depicted in Fig.~\ref{fig:weakswap}. An unrotated PSDP can be used to implement a C-NOT, $\mathds{1}\oplus \sigma_x$, independent of the choice of $|\ell|$ which determines the OAM subspace. 
It is clear from Eq.~\eqref{Eq:PSDP} that the controlled unitary $\mathds{1} \oplus P_{\lambda}^\dagger$ can be implemented using two PSDP's. One of these PSDP's should be rotated by $\alpha = \tfrac{\lambda}{2l}$ and mounted coaxially with the second unrotated PSDP. In order to reduce the number of half- and quarter-wave plates required, we combine the action of one of the rotated half-wave plates from the rotated PSDP with the action of $P^\dagger_{\frac{\lambda}{2}} H$ (App.~\ref{Ap:3}). 

By selecting an appropriate angle of rotation $\alpha$, namely $\alpha=\frac{\pi}{4\ell}$, the optical scheme shown in Fig.~\ref{fig:weakswap} allows us to perform swap operation $S$ between polarisation and a particular OAM subspace, in one iteration. We note that the swap will also be performed for the two-level OAM subspace associated with $\ell'=(4n+1)\ell$, $n\in \mathbb{Z}$, since $2\alpha l' = \frac{\pi}{2}+2\pi n$.

\subsection{Target state dependent control mechanism}\label{Sec:Example_2}
In~\citep{konrad2020robust} it is shown that the unitary 
\begin{align}\label{bla}
U(\lambda) = \exp\left(-i\frac{\lambda}{2} (\sigma_y \otimes \sigma_y + \sigma_z \otimes \sigma_z)\right)\,,
\end{align}
with coupling parameter $\lambda \in \mathbb{R}$ leads to convergence to the target state $\ket{T}=(\ket{0}+\ket{1})/\sqrt{2}$ encoded in the controller system, since the Kraus operators $\{K_i = \braket{i|U|T}\}_i$ satisfy condition~\eqref{eq:fp_contition} and \eqref{eq:span_contition}.  A different target state can be reached by choosing different combinations of Pauli operators as generators.

The following decomposition represents the optical implementation of this unitary~\eqref{bla}:
\begin{align}
U(\lambda)
& = \begin{pmatrix}
  \mathds{1}& 0\\ 0&\sigma_x
\end{pmatrix} 
( H  \otimes \mathds{1} ) 
 \begin{pmatrix} \mathds{1} & 0\\  0& P_{\lambda}^\dagger \end{pmatrix}
( H \otimes  \mathds{1} )
\begin{pmatrix}
  \mathds{1}& 0\\ 0&\sigma_x
\end{pmatrix}\,.\label{Eq:example_2}
\end{align}
The optical implementation for polarisation (controller) and a two-dimensional subspace of OAM spanned by $\{\ket{-\ell},\ket{\ell}\}$ (system) is shown in Fig.~\ref{fig:example2}. 

The unitary in Eq.~\eqref{Eq:example_2} is almost identical to the weak swap discussed in the previous example, Eq.~\eqref{Eq:swap_dec}. However, as a consequence of the absence of phase shift $P_{\frac{\lambda}{2}}^\dagger$ acting on polarisation, a fixed optical implementation achieves $U(\lambda)$ of varying $\lambda$ values depending on the OAM subspace considered.  If we fix the angle $\alpha$ then the apparatus implements the unitary $U(2\alpha \ell)$ on the subspace spanned by $\{\ket{-\ell},\ket{\ell}\}$. 

Consequently, the target state can be reached in all subspaces with close to unit fidelity, if a sufficient number of iterations are performed. This fails if $\alpha \ell$ is equal to an integer multiple of $\pi$. 

For a chosen optical set-up, i.e.\ for  fixed $\alpha$, the fidelity, $F^\ell_n$, for a particular $\ell$ subspace after $n$ iterations is given by the overlap between the final state of the system and the target state. The fidelity increases in an exponential fashion~\cite{konrad2020robust}
\begin{align}
F^\ell_n = 1-(1-F^\ell_0)(1-\sin^2(\alpha \ell))^n\, ,
\end{align}
where $F^\ell_0$ is the initial overlap between the system and target state.  From this we can also determine the number of iterations necessary to reach a certain fidelity $F$
\begin{align}
n = \frac{1}{1-\sin^2(\alpha \ell)} \ln\left(\frac{1-F}{1-F_0^\ell} \right)\,.
\end{align}

\section{Discussion}\label{discussion}

In conclusion, we have presented a class of schemes to prepare the OAM and polarisation qubits using coherent feedback control. Our results are valid for single photons as well as classical beams of light and require mostly linear optical setups. The biggest obstacle in realizing the coherent feedback control in optical systems is to perform the non-local unitaries jointly on the system and controller. This was accomplished by using the CS-decomposition, which reduces the joint unitaries into simple unitaries acting on individual degrees of freedom of light. 
Our coherent feedback control methods allow to prepare arbitrary superpositions of two OAM modes, even without spatial interferometers, and in principle without photon losses. This is an important step forward compared to other preparation methods, such as using spatial light modulators. A generalisation to the preparation of arbitrary structured light modes seems possible.

While for massive systems, measurement-based feedback can be used to prepare target states without losses, non-destructive and efficient measurements are not easily available for photons. Our results show how to translate measurement based feedback into coherent feedback coupling various degrees of freedom of light. This recipe might also be employed for composite massive systems.

Most of the schemes discussed here enable the preparation of a desired state in a single shot. However, coherent feedback control requires in general  iterative interaction of the system and controller. This is for example the case to protect a system against noise or for steering the system into target dynamics. For this purpose, the controller needs to be reset to its initial state, or we need a fresh controller after each iteration. While readily available for systems with strong coupling, for optical systems the situation is more severe.

In optical systems, resetting the controller or using fresh controller leads to exponential increase in the resources. To overcome this problem we have suggested two methods, one involving coherent amplification of light and the other using an additional degree of freedom. Both the methods have their own limitations which result in low fidelities and inefficient implementation of coherent feedback control. However, using these techniques the resources required are linear with the number of interactions.

The coherent feedback control methods discussed here can in principle compensate weak control by repeated application of the control channel, for example the weak swap or the state dependent control scheme. This compensation mechanism is important for photons that weakly couple to their environment and might lead to further applications in quantum communication tasks.   
 
\section{Acknowledgements}
SKG acknowledges the financial support from SERB-DST (File No. ECR/2017/002404). AR and TP acknowledges the financial support of the National Research Foundation of South Africa. We thank G.P. Teja, Akshay Menon and Benjamin Perez-Garcia for useful discussions.

\appendix
\section{Decomposition of the Sine-Cosine matrix}\label{Ap:1}
In the section we provide details of the derivation of Eq.~\eqref{eq:bs_internal_decomp}.
\begin{align}
\mathcal{S} & = \begin{pmatrix} C & S\\ -S & C \end{pmatrix}\\
&= \mathds{1} \otimes C + i \sigma_y \otimes S\\
& = \mathds{1} \otimes \left( \frac{\Theta + \Theta^\dagger}{2} \right) + \sigma_y \otimes \left( \frac{\Theta - \Theta^\dagger}{2} \right)\\
& = \left( \frac{\mathds{1} + \sigma_y }{2}\right) \otimes \Theta  +  \left( \frac{\mathds{1} - \sigma_y }{2} \right) \otimes \Theta^\dagger 
\end{align}
It is straightforward to show that 
\begin{align}
 \frac{\mathds{1} + \sigma_y }{2} & =P_{\tfrac{\pi}{4}}^\dagger H \ket{0}\bra{0} HP_{\tfrac{\pi}{4}} 
 \end{align}
and
\begin{align}
 \frac{\mathds{1} - \sigma_y }{2} & = P_{\tfrac{\pi}{4}}^\dagger H \ket{1}\bra{1}   H P_{\tfrac{\pi}{4}} \,,
\end{align}
so that 
\begin{align}
\mathcal{S} & = \left(P_{\tfrac{\pi}{4}}^\dagger H \otimes\mathds{1}\right)
 \begin{pmatrix}
\Theta & 0\\
0 & \Theta^\dagger
\end{pmatrix}
\left( H P_{\tfrac{\pi}{4}} \otimes\mathds{1}\right)\,.
\end{align}

\section{Decomposition of controlled unitary operations into PSDPs}\label{Ap:2}
We wish to implement an arbitrary controlled unitary of the form
\begin{align}
C_{\Theta} =  \mathds{1} \oplus \begin{pmatrix}
e^{i\theta_1} & 0\\ 0 & e^{i\theta_2}\\ 
\end{pmatrix}\, .
\end{align}
An appropriate phase shift on the first subsystem symmetrise the controlled operation as follows
\begin{align}
C_{\Theta} &= e^{i\theta'_1} \left(\mathds{1} \oplus P_{\theta'_2} \right)  \left(P_{\theta'_1}^\dagger \otimes \mathds{1}\right)\, ,
\end{align}
where $\theta'_1 = \frac{\theta_1+\theta_2}{4}$ and  $\theta'_2 = \frac{\theta_1-\theta_2}{2}$. Discarding the global phase factor we write the controlled unitary $C_\Theta$ as the product 
\begin{align}
C_{\Theta} & = \left(\mathds{1} \oplus \sigma_x\right) \left(\mathds{1} \oplus \sigma_x P_{\theta'_2} \right)\left(P_{\theta'_1}^\dagger \otimes \mathds{1}\right) 
\,.
\end{align}
The C-NOT operation $\mathds{1} \oplus \sigma_x$ can be implemented using a single  PSDP (not rotated) while the the second controlled operation $\mathds{1} \oplus \sigma_x P_{\theta'_2}$ can be implemented using a PSDP rotated by an angle $\theta'_2$ (Eq.~\eqref{Eq:PSDP}). 

\section{Wave plate configuration}\label{Ap:3}
In this section we give the explicit calculation which provides the decomposition of $H_{\frac{\alpha}{2}}P_{\frac{\lambda}{2}} H$ into quarter- and half-wave plates. 
\begin{align}
P_{\frac{\lambda}{2}} &= e^{i\frac{\lambda}{2}\sigma_z}\\
H&= e^{-i\frac{\pi}{4}\sigma_y }\sigma_z = \sigma_z e^{i\frac{\pi}{4}\sigma_y }\\
H_{\frac{\alpha}{2}} & = e^{-i\alpha\sigma_y} \sigma_z\\
\Rightarrow  H_{\frac{\alpha}{2}}P_{\frac{\lambda}{2}} H &=  e^{-i\alpha\sigma_y} e^{i\frac{\lambda}{2}\sigma_z}  e^{i\frac{\pi}{4}\sigma_y }
\end{align}
From Eq.~\eqref{eq:QQH} we have:
\begin{align}
H_{\frac{\alpha}{2}}P_{\frac{\lambda}{2}} H &=  Q_{\alpha+\frac{\pi}{4}} H_{\frac{2\alpha+\lambda}{4}-\frac{\pi}{8}}Q_{\frac{3\pi}{8}}\,.
\end{align}

\section{Examples of iterative coherent feedback using filtering}\label{Ap:fid}

In this section we provided the target fidelity $F_n$ after $n$ iterations of control and filtering for the three examples discussed in Sec.~\ref{Sec:Examples}. 
\begin{align}
F_n &:= \frac{\left|\braket{T|\widetilde{K}^n|\psi_0}\right|^2}{\braket{\psi_0|(\widetilde{K}^\dagger)^n\widetilde{K}^n|\psi_0}}\,,
\end{align}
where $\widetilde{K} := ({K_0+K_1})/{\sqrt{2}}$, $\ket{\psi_0}$ is the initial state of the system and $\braket{T|T_\perp} =0$.
\subsection{Basic control scheme}
In this example the Kraus operators are given by
\begin{align}
K_0& = \ket{T}\bra{T} \\
K_1& =\ket{T}\bra{T_\perp}
\end{align}
so that
\begin{align}
\widetilde{K}  & = \frac{1}{\sqrt{2}}\left(\ket{T}\bra{T} + \ket{T}\bra{T_\perp}\right)\,.
\end{align}
Taken to the $n$th power
\begin{align}
\widetilde{K}^n = \frac{1}{2^{\frac{n}{2}}}\left(\ket{T}\bra{T} + \ket{T}\bra{T_\perp}\right)\,.
\end{align}
Provide that $\braket{T|\psi_0}+\braket{T_\perp|\psi_0} \neq 0$, the target fidelity after each iteration can be shown to be unity, which is to be expected since this scheme works in a single round. 

\subsection{Weak swap}
The Kraus operators associated with the weak swap are given by~\cite{konrad2020robust}
\begin{align}
K_0& = e^{-i\lambda}\ket{T}\bra{T}+ \cos\lambda\ket{T_\perp}\bra{T_\perp} \\
K_1& = \sin\lambda \ket{T}\bra{T_\perp}
\end{align}
so that
\begin{align}
\widetilde{K} & =\frac{1}{\sqrt{2}} \left(e^{-i\lambda} \ket{T}\bra{T} + \cos\lambda \ket{T_\perp}\bra{T_\perp}+ \sin\lambda\ket{T}\bra{T_\perp}  \right)\,.
\end{align}
Taken to the $n$th power
\begin{align}
\widetilde{K}^n = \left(\frac{e^{-i\lambda}}{\sqrt{2}}\right)^n \Bigg( \ket{T}&\bra{T}+ e^{in\lambda}\cos^n\lambda \ket{T_\perp}\bra{T_\perp} \nonumber \\ & +\sin\lambda \sum_{k=0}^{n-1} e^{i(k+1)\lambda}\cos^{k}\lambda \ket{T}\bra{T_\perp} \Bigg)
\end{align}
so that
\begin{align}
F_n & = \frac{1}{1 + \Lambda(\cos\lambda)^{2n}}\,,
\end{align}
where
\begin{align}
\Lambda = \frac{|\braket{T_\perp|\psi_0}|^2}{|\braket{T|\psi_0}+e^{i\lambda}\sin\lambda \sum_{k=0}^{n-1} (e^{i\lambda}\cos\lambda)^{k}\braket{T_\perp|\psi_0} |^2}\,.
\end{align}
The target fidelity $F_n$ converges to one for large $n$ provided that $\lambda$ is not an integer multiple of $\pi$ and $\Lambda<\infty$.

\subsection{Target state dependent control mechanism}
The Kraus operators in this example are given by~\cite{konrad2020robust}
\begin{align}
K_0& = \frac{1}{\sqrt{2}}\left(\ket{T}\bra{T} +\sin\lambda \ket{T}\bra{T_\perp} +\cos\lambda \ket{T_\perp}\bra{T_\perp}\right)\\
K_1& =\frac{1}{\sqrt{2}}\left(\ket{T}\bra{T} -\sin\lambda \ket{T}\bra{T_\perp} +\cos\lambda \ket{T_\perp}\bra{T_\perp}\right)
\end{align}
so that 
\begin{align}
\widetilde{K}& =\ket{T}\bra{T} +\cos\lambda \ket{T_\perp}\bra{T_\perp}\,.
\end{align}
Taken to the $n$th power
\begin{align}
\widetilde{K}^n =\ket{T}\bra{T} +\cos^n\lambda \ket{T_\perp}\bra{T_\perp}\,.
\end{align}
so that
\begin{align}
F_n 
& = \frac{1}{1 + (\cos\lambda)^{2n}\frac{|\braket{T_\perp|\psi_0}|^2}{|\braket{T|\psi_0}|^2}}\,.
\end{align}
The target fidelity $F_n$ converges to one for large $n$ provided that $\lambda$ is not an integer multiple of $\pi$ and $|\braket{T|\psi_0}|> 0$.

\section{Parametric Amplification}\label{Ap:4}
The process of parametric amplification involves the interaction of three fields, the signal $E_1(z,t)$, the idler $E_2(z,t)$ and the pump field $E_3(z,t)$. We follow the method presented in \cite{opampManzoni}. For the case of monochromatic plane waves, where the pump beam is undepleted during the nonlinear interaction and there is no initial idler field, the signal field evolution along the crystal is given by
\begin{equation}
    \frac{\partial^2 A_1}{\partial z^2} = -i \Delta k\ \frac{\partial A_1}{\partial z} + \Gamma^2 A_1. 
\end{equation}
Here $A_1$ refers to the complex amplitude of the signal field, $\Delta k = k_3 - k_2 -k_1$ is the wave vector mismatch and $\Gamma$ is the coupling constant from the nonlinear wave equations defined as
\begin{equation}
    \Gamma^2 = \frac{4 d_{eff}^2 \omega_1^2 \omega_2^2 }{k_1 k_2 c^4} |A_3|^2. 
\end{equation}
The constant $d_{eff}$ relates to the nonlinear susceptibility of the crystal, $\omega_i$ and $k_i$ refer to the angular frequencies and the wave numbers of the fields in the medium, respectively, here $c$ is the speed of light. Since the beam intensity can be given by $I_i = \tfrac{1}{2}n_i \varepsilon_0 c |A_i|^2$, the signal and idler intensities after the interaction length of the nonlinear crystal are
\begin{equation}
    \begin{aligned}
    I_1(L)&= I_1(0) \left(1+ \left(\tfrac{\Gamma}{g} \text{sinh}(gL)   \right)^2  \right)\\
    I_2(L)&= I_1(0)\tfrac{\omega_2}{\omega_1} \left(\tfrac{\Gamma}{g} \text{sinh}(gL)   \right)^2,
    \end{aligned}
\end{equation}
where $I_1(0)$ is the initial signal field intensity, $\varepsilon_0$ is the electric permittivity and $g$ is a generalised wave number given by 
\begin{equation}
    g=\sqrt{\Gamma^2-\frac{\Delta k ^2}{4}}.
\end{equation}
The parametric gain $G$ is therefore defined as the ratio of the signal intensity before and after the nonlinear process, $G=\frac{I_1(L)}{I_1(0)}$.

\section{Creating distinct time-bins} \label{unique}
If the number of round trips in the delay loop (Fig.~\ref{fig:time}) is doubled in each iteration, the delay time $T_N$ a specific pulse spends in the delay loop in $N$ iterations reads
\begin{equation}
    T_{N}(\mathbf{a})= \sum_{n=1}^N a_n\  2^{n}\ \tau \,.
\end{equation}
Here the vector $\mathbf{a}= (a_1, a_2\ldots a_N)$ contains information about which path the pulse took in each iteration. The component $a_i$ is $0$ or $1$, depending on whether the pulse entered the delay loop in the $i$th interation.  The time period $\tau$ is the duration of a single round-trip in the delay loop.  
Since each vector $\mathbf{a}$ is the binary representation of a specific number $T_{N}/\tau$, the corresponding delay $T_{N}$  is  unique and hence the time bins do not overlap.


%

\end{document}